\documentclass[a4paper]{jpconf}
\usepackage{graphicx}
\usepackage{amssymb}
\usepackage{color} 
\usepackage{soul}

\begin{document}
\title{Neutrino parameters and the $N_2$-dominated scenario of leptogenesis\footnote{Contribution to the Proceedings of the NuPhys2013 Conference, 19-20 December 2013, IOP, London.}}

\author{M Re Fiorentin$^a$ and SE King$^{a,b}$}

\address{$^a$ School of Physics and Astronomy, University of Southampton, Southampton, SO17~1BJ, UK}
\address{$^b$ School of Physics and Astronomy, Queen Mary University of London, London, E1~4NS, UK}

\ead{m.re-fiorentin@soton.ac.uk, soph.e.king123@gmail.com}

\begin{abstract}
We briefly review the main aspects of leptogenesis, describing both the unflavoured and the flavoured versions of the $N_2$-dominated scenario. A study of the success rates of both classes of models has been carried out. We comment on these results and discuss corrective effects to this simplest scenario. Focusing on the flavoured case, we consider the conditions required by strong thermal leptogenesis, where the final asymmetry is fully independent of the initial conditions. Barring strong cancellations in the seesaw formula and in the flavoured decay parameters, we show that strong thermal leptogenesis favours a lightest neutrino mass $m_1\gtrsim 10$~meV for normal ordering (NO) and $m_1\gtrsim 3$~meV for inverted ordering (IO). Finally, we briefly comment on the power of absolute neutrino mass scale experiments to either support or severely corner strong thermal leptogenesis.
\end{abstract}

\section{Leptogenesis and the $N_2$-dominated scenario}
Leptogenesis is a particularly attractive process for producing the baryon asymmetry of the Universe, since it utilises the same mechanism that is able to explain the observed neutrino masses and mixing, namely the seesaw mechanism. Considering type-I seesaw, the Standard Model is extended by including heavy right-handed (RH) Majorana neutrinos, $N_i$, that couple to the lepton doublets via Yukawa interactions. Their $CP$- and lepton-number-violating decay produces a lepton asymmetry, which is partly converted into the baryon sector by means of sphaleron processes. We assume in general the presence of 3 RH neutrinos, thus giving rise to a model that depends on 18 parameters.  These split into the 9 so-called {\itshape low-energy neutrino parameters,} probed in experiments, and a further 9 describing the high-energy scale of the RH neutrinos. We relate these two sets of parameters via the complex orthogonal matrix $\Omega$ parameterisation \cite{casas_ibarra}. By definition, {\itshape $N_2$-dominated}~\cite{n2_dominated} scenarios produce the observed asymmetry by the decay of the next-to-lightest RH neutrino, $N_2$.  The subsequent decay of the lightest RH neutrino, $N_1$, produces a negligible contribution and, furthermore, acts to wash the asymmetry out.  These models are particularly interesting because they naturally arise when conditions inspired by $SO(10)$ grand unification models are considered. Indeed, in this framework the spectrum of the RH neutrinos is hierarchical with $M_3>10^{12}$~GeV, $10^9\mbox{ GeV}<M_2<10^{12}\mbox{ GeV}$ and $M_1<10^9$~GeV, hence leptogenesis is $N_2$-dominated. In the unflavoured case, flavour interactions are neglected and RH neutrinos decay into a coherent superposition of flavour eigenstates. The final value of the asymmetry is exponentially suppressed by the unflavoured $N_1$ decay parameter, $K_1\equiv\Gamma_1/H(T=M_1)$, where $H$ is the Hubble parameter and $\Gamma_1$ is the $N_1$ total decay width.  Only when $K_1\ll1$ is the model able to produce the observed asymmetry. A numerical analysis shows that this is realised in only $\sim0.2\%$ of the parameter space.

This picture improves by taking account of flavour interactions. Considering that $\tau$-interactions are efficient for $T\lesssim 10^{12}$~GeV, and muonic ones for $T\lesssim 10^9$~GeV, $N_2$ decays in a two fully-flavoured regime $10^9\mbox{ GeV}\lesssim T\lesssim 10^{12}\mbox{ GeV}$, while the wash-out due to $N_1$ takes place in a three fully-flavoured regime $T\lesssim 10^9$~GeV. Around the transition temperatures the behaviour cannot be described by the usual Boltzmann equations and a density matrix formalism must be adopted \cite{density}. Avoiding these regions, the asymmetry produced by $N_2$ is projected onto the flavour basis and the $N_1$ wash-out acts separately on each flavour $\alpha$ ($\alpha=e,\,\mu,\,\tau$), exponentially suppressing the asymmetry by the relative flavoured decay parameter, $K_{1\alpha}\equiv(\Gamma_{i\alpha}+\bar{\Gamma}_{i\alpha})/H(T=M_1)$. This only requires one $K_{1\alpha}\ll 1$ in order to have a successful model. This is found to happen in $\sim30\%$ of the parameter space, showing that thanks to flavour effects the $N_2$-dominated scenario can represent a viable model of leptogenesis. 

Some approximations are made: in particular, the presence of phantom terms and flavour coupling is neglected \cite{fuller}. The phantom terms correct the projection onto the flavour basis of the asymmetry produced by $N_2$ in the coherent superposition of $e$ and $\mu$ flavours. Each flavour asymmetry can be regarded as a combination of a term proportional to the total lepton asymmetry and a term related to the different flavour composition of the lepton quantum states with respect to their $CP$ conjugate. This contribution is commonly referred to as a {\itshape phantom term}. A further effect, {\itshape flavour coupling}, introduces a correction when one considers that the Boltzmann equations are coupled due to the asymmetries stored in the lepton doublets and in the Higgs bosons. While the size of these effects has already been studied, we focus our attention on the size of the parameter space region where these corrections become significant.  A numerical analysis reveals that for $\sim20\%$ of the parameter space the phantom terms and the flavour coupling modify the final asymmetry by at least one order of magnitude.
\section{Strong thermal leptogenesis}
The final baryon asymmetry depends in general on the initial conditions. At the high temperatures required by this minimal scenario, it is possible that prior to leptogenesis some other mechanisms may generate a significant asymmetry. Leptogenesis models able to wash out a pre-existing asymmetry $N_{B-L}^{p,i}$, while producing the correct final amount, are said to satisfy the {\itshape strong thermal leptogenesis} condition \cite{stl}. It has been shown \cite{stl} that this case is realised only in a two-stage process where the RH neutrinos adhere to the pattern discussed above, thus reinforcing the interest in $N_2$-dominated models. In this scenario, $N_2$ decays in the two-flavoured regime and efficiently washes out the $\tau$ component of $N_{B-L}^{p,i}$. $N_1$, on the other hand, decays in the three-flavoured regime, separately erasing the $e$ and $\mu$ components of $N_{B-L}^{p,i}$, while the produced asymmetry survives in the $\tau$ flavour. This translates into a set of conditions on the relevant decay parameters, namely $K_{1e}, K_{1\mu}, K_{2\tau}\gg1$ and $K_{1\tau}\lesssim1$. When the low-energy neutrino parameters are constrained within the current experimental ranges, our analysis shows that, in the NO case, the strong thermal conditions can be simultaneously satisfied, without fine-tuning, only if the absolute neutrino mass scale $m_1$ is sufficiently large. In particular, for small values of $m_1$ it becomes difficult to obtain a value of $K_{1e}$ large enough to ensure the wash-out of a pre-existing asymmetry, while keeping $K_{1\tau}$ small. It is then possible to place a precise analytical lower bound on $m_1$ \cite{art1}. This lower bound depends on the magnitude of $N_{B-L}^{p,i}$ and on the size of the entries of the orthogonal matrix $\Omega$, i.e. on $\mbox{max}[\left|\Omega_{ij}\right|^2]\equiv M_\Omega$. The larger $M_\Omega$, the more the seesaw mechanism relies, in order to work, on fine-tuned cancellations, rather than on the natural interplay between the electroweak scale and the higher, beyond the Standard Model, scale of the RH neutrinos. 
In general, the lower bound shows a characteristic dependence on the Dirac phase, $\delta$, and has its minimum for $\delta=0$. For $M_\Omega\leq2$ and a large pre-existing asymmetry, $N_{B-L}^{p,i}=0.1$, at $\delta=0$ one gets $m_1\geq0.7$~meV at $95\%$ C.L. The lower bound is relaxed for lower values of $N_{B-L}^{p,i}$ and larger values of $M_\Omega$. In particular, for $M_\Omega\gtrsim 4$ the lower bound disappears. Our results are fully supported by a numerical analysis that demonstrates the analytical lower bound being saturated at the expense of a high level of fine-tuning in the seesaw formula and in the flavoured decay parameters. Moreover, a statistical study of the $m_1$ distribution shows a clear peak around $m_1\sim m_\textup{\scriptsize atm}\simeq50$~meV and rapid suppression for smaller values. In the aforementioned setup, more than $99\%$ of points are found for $m_1\gtrsim 10$~meV (the value quoted in the abstract).

In the IO case, tension exists between the smallness of $K_{1\tau}$ and a $K_{1\mu}$ that must be large enough to wash-out the pre-existing asymmetry. However, with respect to the NO case, $K_{1\mu}$ is easily large enough, even for small values of $m_1$. More precisely, an analytical lower bound on $m_1$ holds for IO only when $M_\Omega\lesssim 0.9$ and is therefore much looser than for NO. The numerical analysis confirms again the analytical results, but now $99\%$ of the points are found for $m_1\gtrsim 3$~meV (the value quoted in the abstract).

It is important to stress again that these results rely on the need for a $K_{1e}$ ($K_{1\mu}$), in NO (IO), large enough to efficiently wash-out the pre-existing asymmetry, while keeping $K_{1\tau}$ small. This request becomes nontrivial due to the experimental values of the neutrino mixing angles, in particular to the suppression introduced by $\sin\theta_{13}$. It can be shown that, with our usual requirements $N_{B-L}^{p,i}=0.1$ and $M_\Omega\leq2$, there is no lower bound on $m_1$ for $30^\circ\lesssim\theta_{13}\lesssim50^\circ$. The reactor mixing angle happens to be large enough to allow both itself and the Dirac phase to be probed, but not too large to trivially satisfy strong thermal leptogenesis and remove any lower bound on $m_1$.

In the coming years, experimental results of the neutrino mass measurements, especially from cosmology, may have the potential to test models of leptogenesis. A measurement of the sum of the neutrino masses can put a constraint on $m_1$ and, if the uncertainty is further reduced to $\sim10$~meV \cite{exp}, it will become possible to discriminate between vanishing and non-vanishing $m_1$. Recent analyses that combine the data from {\itshape Planck} with the latest growth of structure measurements from the Baryon Oscillation Spectroscopic Survey (BOSS) have been able to give for the sum of the neutrino masses a value $\sum_i m_{\nu_i}=\left(360\pm100\right)$~meV~\cite{boss}, at $3.4\sigma$ significance. Together with the measurements from neutrino oscillation experiments this translates into $m_1=\left(127\pm32\right)$~meV for NO and $m_1=\left(117\pm32\right)$~meV for IO. These values are quite high, but they serve as a strong hint at a non-vanishing absolute neutrino mass scale, which goes in the right direction for strong thermal leptogenesis. On the contrary, if new observation will measure $m_1<1$~meV, a very little portion of the parameter space will let strong thermal leptogenesis survive, at the expense of a high level of fine-tuning. Hence, future cosmological observations have the power to either support or severely corner strong thermal leptogenesis. However, this is highly dependent on the ordering and therefore it is of the utmost importance that, in the next years, neutrino oscillation experiments solve the ambiguity between NO and IO. The former is in general a much more favourable case than IO for a significant test, since it hints at larger deviations from the hierarchical limit ($m_1=0$).

\section{Conclusions}
Strong thermal leptogenesis introduces a theoretical request that is able to constrain the parameter space and provide interesting results on $m_1$. In a reasonable setup, favourable values of the absolute neutrino mass scale are found for NO, $m_1\gtrsim10$~meV and IO, $m_1\gtrsim3$~meV. In NO a further analytical lower bound is found for natural choices of $\Omega$ ($M_\Omega\lesssim 4$), while in IO the analytical threshold is obtained only for $M_\Omega\lesssim 0.9$. 
These constraints on $m_1$ allow the future cosmological observations, together with the determination of the neutrino mass ordering, to test strong thermal leptogenesis. 
We conclude by highlighting that the link between cosmology and neutrino phenomenology can prove to be once again extremely fertile and rich of ideas that may shed more and more light on some of the major puzzles of modern physics. 

\section*{Acknowledgements}
SK acknowledges financial support from the NExT/SEPnet Institute. MRF acknowledges financial support from the STAG Institute. MRF is deeply grateful to Pasquale Di Bari, Stefano Moretti and the School of Physics and Astronomy of the University of Southampton. MRF wishes to thank Alexander J. Stuart and is also profoundly indebted to Fiorenza Donato, Mariaelena Boglione and the Physics Department of the University of Torino for the kind hospitality.

\end{document}